\documentclass[10pt]{wlscirep}

\title{Spin-textures of the Bose-Einstein condensates with three kinds of spin-1 atoms}

\author[1]{Y. Z. He}
\author[2]{Y. M. Liu}
\author[1,*]{C. G. Bao}
\affil[1]{School of Physics, Sun Yat-Sen University, Guangzhou, 510275, P. R. China}
\affil[2]{Department of physics, Shaoguan University, shaoguan, 510205, P. R. China}
\affil[*]{Corresponding author: C.G. Bao, stsbcg@mail.sysu.edu.cn}

\begin{abstract}
We have performed a quantum mechanic calculation (including solving the coupled Gross-Pitaevskii equations to obtain the spatial wave functions, and diagonalizing the spin-dependent Hamiltonian in the spin-space to obtain the total spin state) together with an analytical analysis based on a classical model. Then, according to the relative orientations of the spins $S_A$, $S_B$ and $S_C$ of the three species, the spin-textures of the ground state can be classified into two types. In Type-I the three spins are either parallel or anti-parallel to each others, while in Type-II they point to different directions but remain to be coplanar. Moreover, according to the magnitudes of $S_A$, $S_B$ and $S_C$ the spin-textures can be further classified into four kinds, namely, $p$+$p$+$p$ (all atoms of each species are in singlet-pairs), one species in $f$ (fully polarized) and two species in $q$ (a mixture of polarized atoms and singlet-pairs), two in $f$ and one in $q$, and $f$+$f$+$f$. Other combinations are not allowed. The scopes of the parameters that supports a specific spin-texture have been specified. A number of spin-texture-transitions have been found. For Type-I, the critical values at which a transition takes place are given by simple analytical formulae, therefore these values can be predicted.
\end{abstract}

\begin{document}

\flushbottom
\maketitle

\section*{Introduction}

The study of the multi-species Bose-Einstein condensates (BEC) with atoms having nonzero spin is an attractive topic. For these systems, when the temperature is extremely low (say, lower than $10^{-9}$K), the spatial degrees of freedom are nearly frozen and the spin degrees of freedom play essential roles. Various spin-textures will emerge, and they are found to be sensitive to the very weak spin-dependent forces. Therefore, these systems might be ideal for realizing exquisite control.

When the BEC contains only one kind of $N$ spin-1 atoms, the polar phase ($p$-phase) and the ferromagnetic phase ($f$-phase) have been found in the ground state (g.s.) \cite{ref5,ref6,ref7,ref8,ref9,ref10}. In the $p$-phase the spins of atom are two-by-two coupled to zero to form the singlet pairs ($s$-pair), and the total spin of the condensate $S=0$. In the $f$-phase all the spins are fully polarized, i.e., lying along a common direction, and $S=N$. For 2-species BEC it was found in \cite{ml,ref11,ref12,ref13,ref14,ref15,ref16,ref17,ref18,he1,he2} that there are three types of spin-textures, namely, (i) the $p$+$p$ texture where both species are in $p$-phase; (ii) the $f$//$f$ texture where both species are in $f$-phase, and the two total spins (each for a species) are lying either parallel or antiparallel to each other; and (iii) the $f$//$q$ texture where one in $f$-phase and one in quasi-ferromagnetic phase ($q$-phase, a mixture of aligned spins and $s$-pairs).

The above message from 2-species BEC attracts the exploration on the spin-textures of multi-species BEC. Note that, for 3-species BEC, the three intra-species and three inter-species spin-dependent interactions can be repulsive or attractive. Thus, the spin-textures are expected to be very rich. However, this interesting topic is scarcely studied before. This paper is a primary study on this topic. The aim is to clarify the variety of the spin-textures and the related critical phenomena, and the effects of the intra- and inter-species interactions. We believe that the knowledge extracted from 3-species BEC would be in general useful for understanding the spin-textures of many-body systems with multi-species.

We proceed in the following way:

\begin{itemize}
\item From the experience of 2-species BEC, the spin-textures are seriously affected by the compactness of the spatial wave functions (i.e., $\int\varphi_A^4\mathrm{d}\mathbf{r}$ and $\int\varphi_B^4\mathrm{d}\mathbf{r}$) and the overlap (i.e., $\int\varphi_A^2\varphi_B^2\mathrm{d}\mathbf{r}$). For 3-species BEC, $\int\varphi_J^4\mathrm{d}\mathbf{r}$ ($J=A,B,C$) and $\int\varphi_J^2\varphi_{J'}^2\mathrm{d}\mathbf{r}$ are believed to be also important. Therefore, we solve the coupled Gross-Pitaevskii equations (CGP) under the Thomas-Fermi approximation (TFA) to obtain the spatial wave functions. It is well known that the TFA cannot correctly describe the tails of the wave functions. However, when the particle numbers are huge, the gross feature given by the TFA is good. Since only the gross feature is concerned, the TFA is acceptable.

\item Let $S_J$ be the total spin of the $J$-species. When the singlet-pairing force has been neglected, the three $\{S_J\}$ together with the total spin $S$ of the mixture are good quantum numbers is the total spin-states $\Xi$. $\Xi$ is obtained via a diagonalization of the Hamiltonian in the spin-space. In order to extract physical features from $\Xi$, in addition to the good quantum numbers, the averaged angles $\bar{\theta}_{JJ'}$ between $S_J$ and $S_{J'}$ have also been calculated. Thereby various types of spin-textures specified by $\{S_J\}$ and $\{\bar{\theta}_{JJ'}\}$ can be identified. and the transitions among them are found.

\item In addition to the above quantum mechanic (QM) calculation, a corresponding classical model has been proposed and solved analytically. The results from the model are checked via a comparison with those from QM calculation. This model helps greatly to understand the complicated 3-species spin-textures
\end{itemize}

\section*{Hamiltonian and the ground state}

We consider that the condensate is a mixture of three kinds of spin-1 atoms with particle numbers $N_J$ ($J=A$, $B$ or $C$), and they are trapped by isotropic and harmonic potentials $\frac{1}{2}m_Jw_J^2r^2$. The intra-species interaction is $V_J=\sum_{1\leq i<j\leq N_J}\delta(\mathbf{r}_i-\mathbf{r}_j)(c_{J0}+c_{J2}\mathbf{F}_i^J\cdot\mathbf{F}_j^J)$, where $\mathbf{F}_i^J$ is the spin operator of the $i$-th atom of the $J$-species. The inter-species interaction is $V_{JJ'}=\sum_{1\leq i\leq N_J}\sum_{1\leq j\leq N_{J'}}\delta(\mathbf{r}_i-\mathbf{r}_j)(c_{JJ'0}+c_{JJ'2}\mathbf{F}_i^J\cdot\mathbf{F}_j^{J'})$. We introduce two quantities $m$ and $\omega$, and use $\hbar\omega$ and $\lambda\equiv\sqrt{\hbar/(m\omega)}$ as the units for energy and length. Then, the total Hamiltonian is
\begin{equation}
 H
  =  \sum_J
     (\hat{K}_J+V_J)
    +\sum_{J<J'}
     V_{JJ'},
 \label{h}
\end{equation}
where $\hat{K}_J=\sum_{i=1}^{N_J}$ $\hat{h}_J(i)$, $\hat{h}_J(i)=\frac{1}{2}(-\frac{m}{m_J}\nabla_i^2+\gamma_Jr_i^2)$ and $\gamma_J=\frac{m_J\omega_J^2}{m\omega^2}$.

Note that, in the ground state (g.s.), every particles of a kind will condense to a spatial state (say, $\varphi_J$) which is most favorable for binding. Let $\Xi$ denotes a normalized total spin-state. Then the g.s. can be in general written as
\begin{equation}
 \Psi_\mathrm{o}
  =  \prod_{i=1}^{N_A}
     \varphi_A(\mathbf{r}_i)
     \prod_{j=1}^{N_B}
     \varphi_B(\mathbf{r}_j)
     \prod_{k=1}^{N_C}
     \varphi_C(\mathbf{r}_k)
     \Xi.
 \label{twf}
\end{equation}

Let $\vartheta_{S_JM_J}^{N_J}$ denote a normalized and all-symmetric spin-state for the $J$-species where the spins are coupled to $S_J$ and its $Z$-component $M_J$. According to the theory given in \cite{katr}, $N_J-S_J$ must be even, the multiplicity of $\vartheta_{S_JM_J}^{N_J}$ is one (i.e., $\vartheta_{S_JM_J}^{N_J}$ is unique when $S_J$ and $M_J$ are fixed), and the set $\{\vartheta_{S_JM_J}^{N_J}\}$ is complete for all-symmetric spin-states. Let $(\vartheta_{S_A}^{N_A}\vartheta_{S_B}^{N_B})_{S_{AB}M_{AB}}\equiv(S_AS_B)_{S_{AB}M_{AB}}$ be a combined spin-state of the $A$ and $B$-species, in which $S_A$ and $S_B$ are coupled to $S_{AB}$ and $M_{AB}$. Let $((\vartheta_{S_A}^{N_A}\vartheta_{S_B}^{N_B})_{S_{AB}}\vartheta_{S_C}^{N_C})_{SM}\equiv((S_AS_B)_{S_{AB}}S_C)_{SM}$ be a total spin-state of the mixture, in which $S_{AB}$ and $S_C$ are coupled to $S$ and $M$. When the Hamiltonian is given as above, it turns out that $S_A$, $S_B$, $S_C$, $S$ and $M$ are good quantum numbers, but $S_{AB}$ is not. Nonetheless, the states $((S_AS_B)_{S_{AB}}S_C)_{SM}$ form a complete set so that $\Xi$ can be expanded by them.

\subsection*{The coupled Gross-Pitaevskii equations and the spatial wave functions}

For the Hamiltonian given in Eq.(\ref{h}), based on a standard variational approach we can obtain the set of CGP equations for $\varphi_A$ to $\varphi_C$ as \cite{ml}
\begin{eqnarray}
 ( \hat{h}_A
  +\alpha_{AA}\varphi_A^2
  +\alpha_{AB}\varphi_B^2
  +\alpha_{CA}\varphi_C^2
  -\varepsilon_A )
 \varphi_A &=& 0  \label{cgp1} \\
 ( \hat{h}_B
  +\alpha_{AB}\varphi_A^2
  +\alpha_{BB}\varphi_B^2
  +\alpha_{BC}\varphi_C^2
  -\varepsilon_B )
 \varphi_B &=& 0  \label{cgp2} \\
 ( \hat{h}_C
  +\alpha_{CA}\varphi_A^2
  +\alpha_{BC}\varphi_B^2
  +\alpha_{CC}\varphi_C^2
  -\varepsilon_C )
 \varphi_C &=& 0  \label{cgp3}
\end{eqnarray}
where $\varphi_A$, $\varphi_B$ and $\varphi_C$ are required to be normalized.

Since the spin-dependent forces are in general two order weaker than the central forces, as a reasonable approximation, the contribution of the former on the set $\{\alpha_{JJ'}\}$ can be neglected. Then, we have $\alpha_{JJ'}=c_{J0}N_J$ (if $J=J'$) or $\alpha_{JJ'}=c_{JJ'0}N_{J'}$ (if $J\neq J'$).

Since the kinetic energy increases linearly with particle number $N$, while the interaction energy increases with $N^2$, the relative importance of the kinetic terms is very weak when $N$ is very large. In this case, the TFA is a reasonable approximation \cite{polo,he,sr1}. By neglecting the kinetic terms, in a domain where all the $\varphi_J$ are nonzero, the CGP can be written in a matrix form as
\begin{eqnarray}
 \mathfrak{M}\left(
 \begin{array}{c}
   \varphi_A^2 \\
   \varphi_B^2 \\
   \varphi_C^2
 \end{array}
 \right) =\left(
 \begin{array}{c}
   \varepsilon_A-\gamma_Ar^2/2 \\
   \varepsilon_B-\gamma_Br^2/2 \\
   \varepsilon_C-\gamma_Cr^2/2
 \end{array}
 \right),
 \label{mat}
\end{eqnarray}
where $\mathfrak{M}$ is a $3\times 3$ matrix with elements $\alpha_{JJ'}$. Let the determinant of $\mathfrak{M}$ be $\mathfrak{D}$. From the above matrix equation, we obtain a formal solution of the CGP as
\begin{eqnarray}
 &\varphi_J^2
  =  Z_J
    -Y_J
     r^2,\ \ \
     (J=A,B,C)
 \label{eq5}& \\
 &Z_J
  =  \mathfrak{D}_J^Z/
     \mathfrak{D}.
 \label{eq6}&
\end{eqnarray}
$\mathfrak{D}_J^Z$ is a determinant obtained by changing the $J$ column of $\mathfrak{D}$ from $(\alpha_{AJ},\alpha_{BJ},\alpha_{CJ})$ to $(\varepsilon_A,\varepsilon_B,\varepsilon_C)$.
\begin{equation}
 Y_J
  =  \mathfrak{D}_J^Y/
     \mathfrak{D}.
 \label{eq7}
\end{equation}
$\mathfrak{D}_J^Y$ is also a determinant obtained by changing the $J$ column of $\mathfrak{D}$ to $(\gamma_A/2,\gamma_B/2,\gamma_C/2)$. Once all the parameters are given, the three $Y_J$ are known because they depend only on $\alpha_{JJ'}$ and $\gamma_J$. However, the three $Z_J$ have not yet been known because they depend on $(\varepsilon_A,\varepsilon_B,\varepsilon_C)$. When $Y_J$ is positive (negative), $\varphi_J^2$ goes down (up) with $r$. Thus the main feature of this formal solution depends on the signs of the set $\{Y_J\}$.

The set $\{Z_J\}$ and the set $\{\varepsilon_J\}$ are related as
\begin{eqnarray}
 \varepsilon_J
 &=& \sum_{J'}
     \alpha_{JJ'}
     Z_{J'}, \label{eq7a} \\
 Z_J
 &=& \sum_{J'}
     \bar{\alpha}_{JJ'}
     \varepsilon_{J'}. \label{eq7b}
\end{eqnarray}
where $\bar{\alpha}_{JJ'}=\mathfrak{d}_{J'J}/\mathfrak{D}$, and $\mathfrak{d}_{J'J}$ is the algebraic cominor of $\alpha_{J'J}$. This formal solution is named the Form III, which is valid only in a domain where all the three $\varphi_J$ are nonzero.

When two wave functions are nonzero inside a domain while the third is zero, in a similar way we obtain
\begin{eqnarray}
 \left\{
 \begin{array}{l}
   \varphi_l^2 = Z_l^{(n)}-Y_l^{(n)}r^2  \\
   \varphi_m^2 = Z_m^{(n)}-Y_m^{(n)}r^2  \\
   \varphi_n^2 = 0
 \end{array}
 \right., \label{eq8}
\end{eqnarray}
where $l$, $m$ and $n$ are a cyclic permutation of $A$, $B$ and $C$.
\begin{eqnarray}
 \left\{
 \begin{array}{l}
   Z_l^{(n)} = (\alpha_{mm}\varepsilon_l-\alpha_{lm}\varepsilon_m)/\mathfrak{d}_{nn} \\
   Y_l^{(n)} = \frac{1}{2}(\alpha_{mm}-\alpha_{lm})/\mathfrak{d}_{nn} \\
   Z_m^{(n)} = (\alpha_{ll}\varepsilon_m-\alpha_{ml}\varepsilon_l)/\mathfrak{d}_{nn} \\
   Y_m^{(n)} = \frac{1}{2}(\alpha_{ll}-\alpha_{ml})/\mathfrak{d}_{nn}
 \end{array}
 \right. \label{eq9}
\end{eqnarray}
Once the parameters are given, the six $Y_{n'}^{(n)}$ ($n'\neq n$) are known, while the six $Z_{n'}^{(n)}$ have not yet. This formal solution with $\varphi_n=0$ is denoted as Form II$_n$, where the subscript specifies the vanishing wave function.

When one and only one of the wave functions is nonzero in a domain (say, $\varphi_J\neq 0$), it must have the unique form as
\begin{equation}
 \varphi_J^2
  =  \frac{1}{\alpha_{JJ}}
     ( \varepsilon_J
      -\gamma_Jr^2/2 ).
 \label{eq10}
\end{equation}
Obviously, $\varphi_J$ in this form must descend with $r$. This form is denoted as Form I$_J$, where the subscript specifies the survived wave function.

If a wave function (say, $\varphi_J$) is nonzero in a domain but becomes zero when $r\geq r_\mathrm{o}$, then a downward form-transition (say, from Form III to II$_J$) will occur at $r_\mathrm{o}$. Whereas if $\varphi_J$ is zero in a domain but emerges from zero when $r\geq r_\mathrm{o}$, then an upward form-transition (say, from Form II$_J$ to III) will occur at $r_\mathrm{o}$. $r_\mathrm{o}$ appears as the boundary separating the two connected domains, each supports a specific form. In this way the formal solutions serve as the building blocks, and they will link up continuously to form an entire solution of the CGP. They must be continuous at the boundary because the two sets of wave functions by the two sides of the boundary satisfy exactly the same set of nonlinear equations at the boundary.

Recall that there are three unknowns $\varepsilon_A$, $\varepsilon_b$ and $\varepsilon_c$ contained in the formal solutions. Taking the three additional equations of normalization $\int\varphi_J^2\mathrm{d}\mathbf{r}=1$ into account, the three unknowns can be obtained. Then, under the TFA, the CGP is completely solved. The details are shown below.

\subsection*{The spatial wave functions}

\begin{figure}[tbp]
 \centering \resizebox{0.6\columnwidth}{!}{\includegraphics{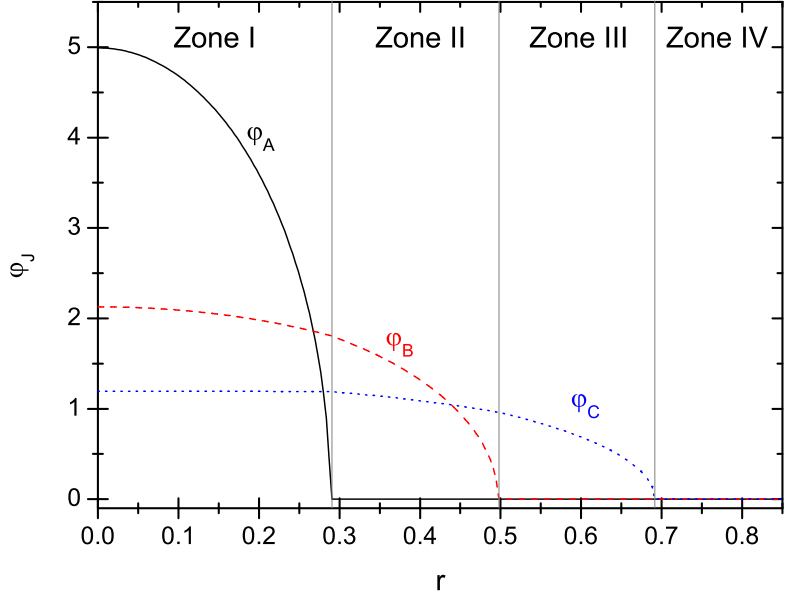} }
 \caption{(color online) An example of the spatial wave functions of a miscible state obtained from the TFA solution of the CGP. The parameters are given as $Y_A=300$, $Y_B=15$, $Y_C=0.1$, $Y^{(A)}_B=20$, $Y^{(A)}_C=3$, $\alpha_{CC}=0.01$ and $\gamma_C=0.08$.}
 \label{fig1}
\end{figure}

Note that the variety of the spin-textures in multi-species BEC is caused by the inter-species interactions. Obviously, they will play a more essential role when the three kinds of atoms are distributed closer to each other. Therefore, in the following examples, we take the miscible states into account, in which all the three species have nonzero distribution at the center ($r=0$). An example is given in Fig.\ref{fig1}, where the wave functions in zone I to IV are in Form III, Form II$_1$, Form I$_3$, and empty, respectively.

For this example, we know that the boundary $r_a$ (at which $\varphi_A=0$) is equal to $\sqrt{Z_A/Y_A}$ (refer to Eq.(\ref{eq5})), $r_b$ (at which $\varphi_B=0$) is equal to $\sqrt{Z_B^{(A)}/Y_B^{(A)}}$ (Eq.(\ref{eq8})), $r_c$ (at which $\varphi_C=0$) is equal to $\sqrt{2\varepsilon_C/\gamma_C}$ (Eq.(\ref{eq10})). They give the outmost boundary of $\varphi_A$, $\varphi_B$ and $\varphi_C$, respectively. Taking the normalization into account, we obtain
\begin{eqnarray}
 Z_A
 &=& (\frac{15}{8\pi})^{2/5}
     Y_A^{3/5}, \label{z1} \\
 Z_B^{(A)}
 &=& (\frac{15}{8\pi})^{2/5}
     (Y_B^{(A)})^{3/5}
     [ 1
      -(Y_B-Y_B^{(A)})
       /Y_A ]^{2/5}, \label{zab} \\
 Z_B
 &=& Z_B^{(A)}
    +(\frac{15}{8\pi})^{2/5}
     (Y_B-Y_B^{(A)})
     /Y_A^{2/5}, \label{zb} \\
 \varepsilon_C/\alpha_{CC}
 &=& (\frac{15}{8\pi})^{2/5}
     (\frac{\gamma_C}{2\alpha_{CC}})^{3/5}
     [ 1
      -\frac{Y_C-Y_C^{(A)}}{Y_A}
      -( Y_C^{(A)}
        -\frac{\gamma_C}{2\alpha_{CC}})
       \frac{1}{Y_B^{(A)}}
       ( 1
        -\frac{Y_B-Y_B^{(A)}}{Y_A} ) ]^{2/5}, \label{z2} \\
 Z_C^{(A)}
 &=& \frac{\varepsilon_C}{\alpha_{CC}}
     +( Y_C^{(A)}
       -\frac{\gamma_C}{2\alpha_{CC}} )
      \frac{Z_B^{(A)}}{Y_B^{(A)}}, \label{zac} \\
 Z_C
 &=& Z_C^{(A)}
    +(Y_C-Y_C^{(A)})
     (\frac{15}{8\pi Y_A})^{2/5}. \label{zc}
\end{eqnarray}
Since $Z_A$, $Z_B$, and $Z_C$ have been obtained as given above, $\varepsilon_A$ and $\varepsilon_B$ can be further obtained via Eq.(\ref{eq7a}). Then, the entire solution of the CGP together with the chemical potentials are completely known.

Nonetheless, the realization of the miscible state is based on a number of assumptions. First, it is assumed that all the wave functions are nonzero at the center, thus $Z_A>0$, $Z_B>0$, and $Z_C>0$ are required. Second, $\varphi_A$ is assumed to descend with $r$ in zone I and $\varphi_B$ is assumed to descend with $r$ in zone II, thus $Y_A>0$ and $Y_B^{(A)}>0$ are required. Third, $\varphi_B|_{r_a}>0$ and $\varphi_C|_{r_a}>0$ are required so that the Form III can link with a Form II$_A$ at $r_a$. Fourth, $\varphi_C|_{r_b}>0$ is required so that the Form II$_A$ can link with a Form I$_C$ at $r_b$. Each of these requirements will impose a constraint on the parameters (say, the requirement $\varphi_B|_{r_a}>0$ leads to $Z_B^{(A)}>Y_B^{(A)}r_a^2$, and therefore leads to $Y_A>Y_B$). Thus, the type as shown in Fig.\ref{fig1} can be realized only if the parameters are given inside a specific scope. A comprehensive discussion on the scope of parameters for each spatial type of solution is the base for obtaining the phase-diagrams, but this is beyond the scope of this paper.

\section*{The total spin-state}

Recall that, in the spin-space, $\Xi$ can be expanded via the basis-states $((S_AS_B)_{S_{AB}}S_C)_{SM}$, where $S_A$, $S_B$, $S_C$, $S$ and $M$ are good quantum numbers, and $S_{AB}$ is ranged from $|S_A-S_B|$ to $S_A+S_B$. Taking the spatial states into account, we define a set of basis-states for the g.s. as
\begin{equation}
 \psi_{\mathfrak{S},S_{AB}}
  =  \prod_{i=1}^{N_A}
     \varphi_A(\mathbf{r}_i)
     \prod_{j=1}^{N_B}
     \varphi_B(\mathbf{r}_j)
     \prod_{k=1}^{N_C}
     \varphi_C(\mathbf{r}_k)
     ((S_AS_B)_{S_{AB}}S_C)_{SM}. \label{twfp}
\end{equation}
where the subscript $\mathfrak{S}$ denotes a specific set $(S_AS_BS_CS) $. When a magnetic field is not applied, the label $M$ can be neglected. Accordingly, a candidate of the g.s. can be expanded as
\begin{equation}
 \Psi_{\mathfrak{S}}
  =  \sum_{S_{AB}}
     d_{S_{AB}}
     \psi_{\mathfrak{S},S_{AB}}, \label{gs}
\end{equation}
Let $H$ be divided as $H=H_{\mathrm{o}}+H_\mathrm{spin}$, where all the spin-dependent interactions are included in $H_\mathrm{spin}$. Let $(J_-,J,J_+)$ be a cyclic permutation of $(A,B,C)$. Then $H_\mathrm{spin}=\sum_Jc_{J2}\sum_{1\leq i<j\leq N_J}\delta(\mathbf{r}_i-\mathbf{r}_j)\mathbf{F}_i^J\cdot\mathbf{F}_j^J+\sum_Jc_{JJ_+2}\sum_{1\leq i\leq N_J}\sum_{1\leq j\leq N_{J_+}}\delta(\mathbf{r}_i-\mathbf{r}_j)\mathbf{F}_i^J\cdot\mathbf{F}_j^{J_+}$. When $\mathfrak{S}$ is given, the coefficients $d_{S_{AB}}$ can be obtained via a diagonalization of $H_\mathrm{spin}$ in the space expanded by $\psi_{\mathfrak{S},S_{AB}}$. The matrix elements are
\begin{eqnarray}
 \langle
 \psi_{\mathfrak{S},S_{AB}'}|
 H_\mathrm{spin}|
 \psi_{\mathfrak{S},S_{AB}}
 \rangle
 &\equiv&
     H_{S_{AB}',S_{AB}}  \nonumber \\
 &=& \delta_{S_{AB}'S_{AB}}
     [ \sum_J\frac{1}{2}
       \int
       \varphi_J^4
       \mathrm{d}\mathbf{r}\
       c_{J2}(T_J-2N_J)
    +\int
     \varphi_A^2
     \varphi_B^2
     \mathrm{d}\mathbf{r}\
     c_{AB2}
     \frac{T_{AB}-T_A-T_B}{2} ]  \nonumber \\
  &&+\int
     \varphi_B^2
     \varphi_C^2
     \mathrm{d}\mathbf{r}\
     c_{BC2}
     \sum_{S_{BC}}
     \bar{w}(S_AS_BSS_C;S_{AB}S_{BC})
     \bar{w}(S_AS_BSS_C;S_{AB}'S_{BC})
     \frac{1}{2}
     (T_{BC}-T_B-T_C)  \nonumber \\
  &&+\int
     \varphi_C^2
     \varphi_A^2
     \mathrm{d}\mathbf{r}\
     c_{CA2}
     \sum_{S_{CA}}
     (-1)^{S_{AB}'+S_{AB}}
     \bar{w}(S_BS_ASS_C;S_{AB}S_{CA})  \nonumber \\
  && \bar{w}(S_BS_ASS_C;S_{AB}'S_{CA})
     \frac{1}{2}
     (T_{CA}-T_C-T_A),  \label{hspinm}
\end{eqnarray}
where the summation of $J$ covers $A$, $B$ and $C$, $\bar{w}(S_AS_BSS_C;S_{AB}S_{BC})=\sqrt{(2S_{AB}+1)(2S_{BC}+1)}w(S_AS_BSS_C;S_{AB}S_{BC})$, the latter is the W-coefficients of Racah, $T_J=S_J(S_J+1)$, and so on.

Carrying out the diagonalization, the lowest eigenstate is $\Psi_{\mathfrak{S}}$ and the corresponding energy is denoted as $E_\mathfrak{S}$. Let the four presumed values in $\mathfrak{S}$ be varied within a reasonable scope. When $\mathfrak{S}=\mathfrak{S}_\mathrm{o}$, if $E_\mathfrak{S}$ arrives at its minimum, then the g.s. $\Psi_\mathrm{o}=\Psi_{\mathfrak{S}_\mathrm{o}}$.

To extract information on spin-texture from $\Psi_\mathrm{o}$, we calculate the averaged angle between the two spins $S_A$ and $S_B$ as
\begin{equation}
 \bar{\theta}_{AB}
  \equiv
     \cos^{-1}[ \langle
                \Psi_\mathrm{o}|
                \hat{\mathbf{S}}_A\cdot
                \hat{\mathbf{S}}_B|
                \Psi_\mathrm{o}
                \rangle/
                \sqrt{ \langle
                       \Psi_\mathrm{o}|
                       \hat{S}_A^2|
                       \Psi_\mathrm{o}
                       \rangle
                       \langle
                       \Psi_\mathrm{o}|
                       \hat{S}_B^2|
                       \Psi_\mathrm{o}
                       \rangle } ]
  =  \cos^{-1}[ \frac{1}{2\sqrt{T_AT_B}}
                \sum_{S_{AB}}
                d_{S_{AB}}^2
                (T_{AB}-T_A-T_B) ], \label{ctab}
\end{equation}
where $\hat{\mathbf{S}}_J\equiv\sum_i\mathbf{F}_i^J$ is the operators for the total spin of the $J$-species. Similarly, we have
\begin{eqnarray}
 \bar{\theta}_{BC}
 &=& \cos^{-1}[\frac{1}{2\sqrt{T_BT_C}}
               \sum_{S_{AB},S_{AB}',S_{BC}}
               d_{S_{AB}}
               d_{S_{AB}'}
               \bar{w}(S_AS_BSS_C;S_{AB}S_{BC})
               \bar{w}(S_AS_BSS_C;S_{AB}'S_{BC})
               (T_{BC}-T_B-T_C)],  \label{ctbc} \\
 \bar{\theta}_{CA}
 &=& \cos^{-1}[\frac{1}{2\sqrt{T_AT_C}}
               \sum_{S_{AB},S_{AB}',S_{CA}}
               d_{S_{AB}}
               d_{S_{AB}'}
               (-1)^{S_{AB}'+S_{AB}}
               \bar{w}(S_BS_ASS_C;S_{AB}S_{CA})
               \bar{w}(S_BS_ASS_C;S_{AB}'S_{CA})
               (T_{CA}-T_C-T_A)].\ \ \  \label{ctca}
\end{eqnarray}
Examples are given below.

\section*{Classical model (Type-I)}

The total energy of the g.s. can be divided as $E=E_\mathrm{o}+E_\mathrm{spin}$, where
\begin{equation}
 E_\mathrm{spin}
  =  \langle
     \Psi_\mathrm{o}|
     H_\mathrm{spin}|
     \Psi_\mathrm{o}
     \rangle
  =  \sum_J
     Q_J
     \langle
     \Xi|
     \hat{S}_J^2-2N_J|
     \Xi
     \rangle
    +2\sum_J
     Q_{JJ_+}
     \langle
     \Xi|
     \hat{\mathbf{S}}_J\cdot
     \hat{\mathbf{S}}_{J_+}|
     \Xi
     \rangle, \label{esp}
\end{equation}
where $Q_J=\int\varphi_J^4\mathrm{d}\mathbf{r}\ c_{J2}/2$, $Q_{JJ_+}=\int\varphi_J^2\varphi_{J_+}^2\mathrm{d}\mathbf{r}\ c_{JJ_+2}/2$.

To see clearer the physical picture, we propose a classical model to facilitate qualitative analysis. In this model, the total spin of the $J$-species is considered as a vector $\vec{S}_J$ with norm $S_J$ ranging from 0 to $N_J$, $\theta_{JJ_+}$ is the angle between $\vec{S}_J$ and $\vec{S}_{J_+}$. The magnitudes and orientations of the three $\vec{S_J}$ together describe an intuitive picture of the spin-texture. The classical analog of $E_\mathrm{spin}$ is defined as
\begin{equation}
 E_\mathrm{spin}^\mathrm{M}
  =  \sum_J
     Q_J
     S_J^2
    +2\sum_J
     Q_{JJ_+}
     S_J
     S_{J_+}
     \cos\theta_{JJ_+}, \label{em}
\end{equation}
The effect of the inter-species force is embodied by $Q_{JJ_+}$. When $Q_{JJ_+}<0$ (attractive), $\vec{S}_J$ and $\vec{S}_{J_+}$ will be lying along the same direction. Whereas when $Q_{JJ_+}>0$ (repulsive), along opposite directions. Note that, for three spins, two of them will define a plane and will pull the third lying on the same plane. Therefore, the spin-textures of 3-species condensates are assumed to be coplanar (this assumption will be checked later). Thus, in what follows, $\theta_{AB}+\theta_{BC}+\theta_{CA}=2\pi$ is given. Accordingly, When $\{Q_J\}$ and $\{Q_{JJ_+}\}$ are given, $E_\mathrm{spin}^\mathrm{M}$ is a function of five variables $(S_A,S_B,S_C,\theta_{BC},\theta_{CA})$. When these variables lead to the minimum of $E_\mathrm{spin}^\mathrm{M}$, they specify a coplanar spin-texture of the g.s.. In order to find out the minimum, we calculate the partial derivatives of $E_\mathrm{spin}^\mathrm{M}$. They are given in the appendix.

\begin{figure}[tbp]
 \centering \resizebox{0.6\columnwidth}{!}{\includegraphics{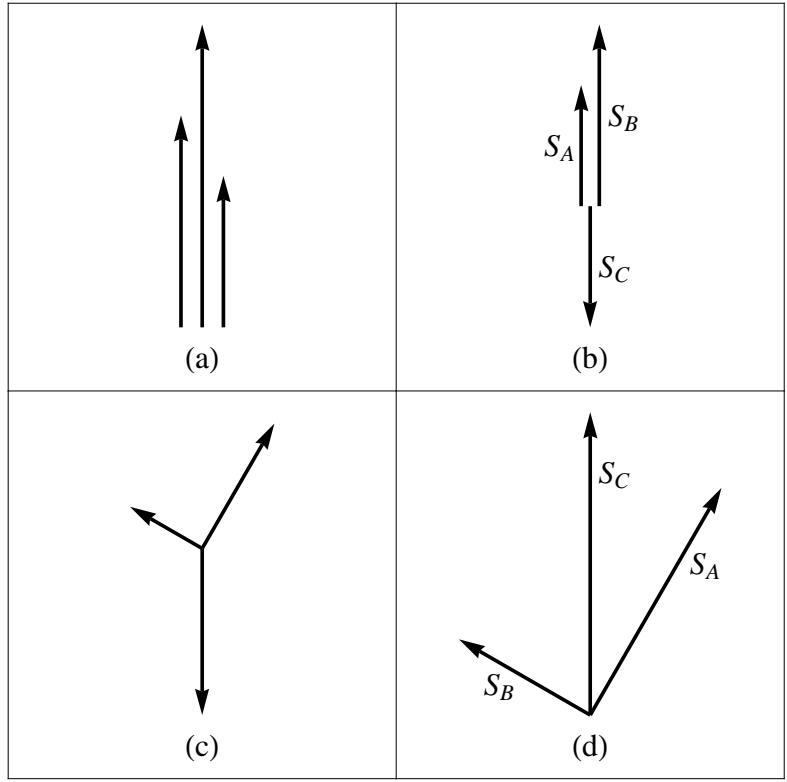} }
 \caption{Intuitive pictures of the coplanar spin-textures, where the relative orientations of the spins $S_A$, $S_B$ and $S_C$ are shown.}
 \label{fig2}
\end{figure}

There are two types of spin-textures. When all $\{Q_{JJ_+}\}$ are negative, $\vec{S}_A$, $\vec{S}_B$ and $\vec{S}_C$ would tend to be parallel to each others, i.e., all $\cos\theta_{JJ_+}=1$ as shown in Fig.\ref{fig2}a. When only one of $\{Q_{JJ_+}\}$ is negative, say, $Q_{AB}$ is negative, orientations of the spins are shown in Fig.\ref{fig2}b, where $\cos\theta_{AB}=1$, $\cos\theta_{BC}=\cos\theta_{CA}=-1$. These two cases are called in Type-I.

For Type-I, the total energy appears as
\begin{equation}
 E_\mathrm{spin}^\mathrm{M}
  =  \sum_J
     ( Q_JS_J^2
      -2|Q_{JJ_+}|S_JS_{J_+} ).
 \label{emt1}
\end{equation}
Let $p$ denotes a point in the 3-dimensional coordinate-space with the coordinates $(S_A,S_B,S_C)$. This point is bound by a cuboid as shown in Fig.\ref{fig3}. When $S_J$ of a species is given, the phase of the species is denoted, for short, by $p$, $f$ and $q$ if $S_J=0$, $N_J$ and in between. Let $p_\mathrm{g.s.}$ be the point where $E_\mathrm{spin}^\mathrm{M}$ arrives at its minimum. There are the following possibilities:

\begin{figure}[tbp]
 \centering \resizebox{0.6\columnwidth}{!}{\includegraphics{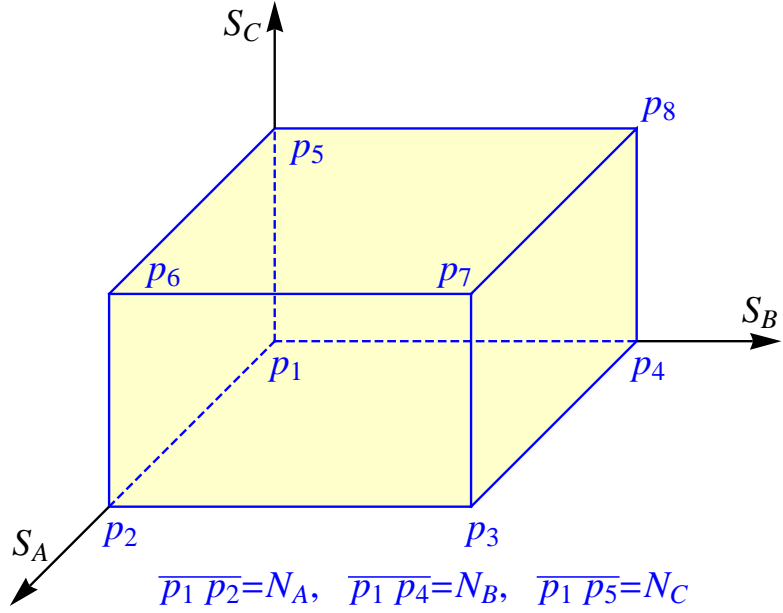} }
 \caption{(color online) The cuboid formed by the norms of the three spins $S_A$, $S_B$, and $S_C$ each from 0 to $N_J$.}
 \label{fig3}
\end{figure}

\subsubsection*{The case $p_\mathrm{g.s.}$ is located inside the cuboid (i.e., not on the surfaces, edges and vertexes).}

In this case $0<S_J<N_J$ for all $J$. At the minimum the three equations $\frac{\partial E_\mathrm{spin}^\mathrm{M}}{\partial S_J}|_{p_\mathrm{g.s.}}=0$ are necessary to hold. This leads to a set of homogeneous linear equations for $(S_A,S_B,S_C)$ as
\begin{equation}
  Q_J
  S_J
 -|Q_{J_-J}|
  S_{J_-}
 -|Q_{JJ_+}|
  S_{J_+}=0,\ \ \
  (J=A,B,C)  \label{em6}
\end{equation}
However, the matrix of this set is in general not singular. Therefore, there is no nonzero solution. Even, for a specific choice of the parameters, the matrix is singular, the nonzero solution can be multiplied by a variable common number $\varsigma$. One can see that $E_\mathrm{spin}^\mathrm{M}$ varies with $\varsigma$ monotonically. In order to minimize $E_\mathrm{spin}^\mathrm{M}$, $\varsigma$ should be given either in its upper or lower limit but not inside. Thus, $p_\mathrm{g.s.}$ cannot locate inside the cuboid. It implies that the three species cannot all in the $q$-phase.

Let a rectangle on the surface of the cuboid be denoted as $p_1p_4p_8p_5$, etc. (refer to Fig.\ref{fig3}). There are six rectangles, three of them contain $p_1$ as a vertex (the first kind), the other three contain $p_7$ as a vertex (the second kind).

\subsubsection*{The case $p_\mathrm{g.s.}$ is located on a rectangle of the first kind.}

There are three such rectangles. If $p_\mathrm{g.s.}$ were located on $p_1p_4p_8p_5$ (i.e., $S_A=0$, $0\leq S_B\leq N_B$, and $0\leq S_C\leq N_C$), it is necessary to have $\frac{\partial E_\mathrm{spin}^\mathrm{M}}{\partial S_A}|_{p_\mathrm{g.s.}}\geq 0$. However, this leads to $-|Q_{AB}|S_B-|Q_{CA}|S_C\geq 0$ which cannot be realized unless $S_B=S_C=0$. With similar arguments, $p_\mathrm{g.s.}$ cannot located on $p_1p_5p_6p_2$ and $p_1p_2p_3p_4$ as well, but it can locate at the point $p_1$. Thus this case is prohibited unless $p_\mathrm{g.s.}=p_1$. It implies that the case with one or two species in $p$-phase is prohibited, while all species in $p$ is possible. This fact coincides with the finding found in 2-species condensates, in which the $p$-phase is extremely fragile when it is accompanied by an $f$ or a $q$. Therefore, the $p$+$f$ or $p$+$q$ textures do not exist, but the $p$+$p$ texture is allowed \cite{ref11,ref12,he1,he2}).

With the above prohibitions, $p_\mathrm{g.s.}$ can only access $p_1$, $p_7$, the interior of the three rectangles of the second kind, and the interior of the three edges $\overline{p_7p_6}$, $\overline{p_7p_3}$, and $\overline{p_7p_8}$.

\subsubsection*{The case $p_\mathrm{g.s.}=p_7$.}

In this case $S_J=N_J$ for all $J$ and, accordingly, the texture is denoted as $f$//$f$//$f$. (the symbol // implies that the related spins are either parallel or anti-parallel). The three inequalities $\frac{\partial E_\mathrm{spin}^\mathrm{M}}{\partial S_J}|_{p_7}<0$ are required to hold. This leads to the constraints imposed on the parameters as listed at the right of the first row of Tab.\ref{tab1}. These constraints give the scope of the parameters that supports the $f$//$f$//$f$-texture. The energy of this texture $E_\mathrm{spin}^\mathrm{M}=E_{fff}^\mathrm{M}$ is listed in Tab.\ref{tab2}. In these tables, we have defined
\begin{equation}
 \alpha_{JJ_+}
  \equiv
     Q_J
     Q_{J_+}
    -|Q_{JJ_+}|^2, \label{abc}
\end{equation}
and
\begin{equation}
 \alpha_{ABC}
  \equiv
     Q_A
     Q_B
     Q_C
    -2|Q_{AB}|
     |Q_{BC}|
     |Q_{CA}|
    -Q_AQ_{BC}^2
    -Q_BQ_{CA}^2
    -Q_CQ_{AB}^2. \label{aabc}
\end{equation}

When every species is ferromagnetic in nature (i.e., all $Q_J<0$), the inequality $N_JQ_J-N_{J_-}|Q_{J_-J}|-N_{J_+}|Q_{JJ_+}|<0$ holds definitely, and the $f$//$f$//$f$ texture is the only choice for the g.s.. When some species (say, $J$-species) is polar in nature (i.e., $Q_J>0$), the term $N_JQ_J$ (representing the intra-interaction) and the other two terms (representing the combined inter-interaction) are competing. Only when $|Q_{J_-J}|$ and $|Q_{JJ_+|}$ are sufficiently large the $J$-species could be in $f$-phase. Note that, in $Q_J$, the strength $c_{J2}$ is weighted by $\int\varphi_J^4\mathrm{d}\mathbf{r}/2$, while in $Q_{JJ'}$, $c_{JJ'2}$ is weighted by $\int\varphi_J^2\varphi_{J'}^2\mathrm{d}\mathbf{r}/2$. Thus, the profiles of the spatial wave functions are important to the spin-textures.

\begin{table}[tbh]
 \caption{When all $Q_{JJ_+}<0$ or only one $Q_{JJ_+}<0$, the representative possible spin-textures of the g.s. are listed in the first column. The notation $f$//$f$//$q$ implies that the $A$, $B$ and $C$ species are in $f$, $f$ and $q$, respectively. The three spins $S_A$, $S_B$ and $S_C$ are either parallel or anti-parallel to each others. The (in)equalities listed in the second column impose a constraint on the parameters so that the associated texture can emerge only in a subspace in the parameter space. In the first row $J=A$, $B$ and $C$. $(J_-,J,J_+)$ is a cyclic permutation of $(A,B,C)$. The constraints for other possible textures not listed in the table, say, $f$//$q$//$f$, can be obtained by a cyclic permutation of the indexes $A$, $B$ and $C$.}
 \label{tab1}
 \begin{center}
 \begin{tabular}{c|c}
 \hline\hline
 spin-texture   & constraint                                            \\ \hline
 $f$//$f$//$f$  & $N_JQ_J-N_{J_-}|Q_{J_-J}|-N_{J_+}|Q_{JJ_+}|<0$        \\ \hline
 $f$//$f$//$q$  & $N_AQ_A-N_B|Q_{AB}|-S_C|Q_{CA}|<0$                    \\
                & $N_BQ_B-S_C|Q_{BC}|-N_A|Q_{AB}|<0$                    \\
                & $S_CQ_C-(N_A|Q_{CA}|+N_B|Q_{BC}|)=0$                  \\
                & $Q_C>0$                                               \\ \hline
 $f$//$q$//$q$  & $N_AQ_A-S_B|Q_{AB}|-S_C|Q_{CA}|<0$                    \\
                & $S_B=N_A(Q_C|Q_{AB}|+|Q_{BC}||Q_{CA}|)/\alpha_{BC}$   \\
                & $S_C=N_A(Q_B|Q_{CA}|+|Q_{BC}||Q_{AB}|)/\alpha_{BC}$   \\
                & $Q_B>0$, $Q_C>0$, $\alpha_{BC}>0$                     \\ \hline
 $p$+$p$+$p$    & $\alpha_{ABC}\geq 0,Q_A>0$, $Q_B>0$, $Q_C>0$          \\
 \hline\hline
 \end{tabular}
 \end{center}
\end{table}

\begin{table}[tbh]
 \caption{The model energies of the g.s. in various textures.}
 \label{tab2}
 \begin{center}
 \begin{tabular}{c|c}
 \hline\hline
 model                  & energy                                                                                        \\ \hline
 $E_{fff}^\mathrm{M}$   & $\sum_J(Q_JN_J^2-2|Q_{JJ_+}|N_JN_{J_+})$                                                      \\
 $E_{ffq}^\mathrm{M}$   & $\frac{1}{Q_C}[N_A^2\alpha_{CA}+N_B^2\alpha_{BC}-2N_AN_B(Q_C|Q_{AB}|+|Q_{BC}||Q_{CA}|)]$      \\
 $E_{fqq}^\mathrm{M}$   & $\frac{N_A^2}{\alpha_{BC}}\alpha_{ABC}$                                                       \\
 $E_{ppp}^\mathrm{M}$   & $0$                                                                                           \\
 \hline\hline
 \end{tabular}
 \end{center}
\end{table}

\subsubsection*{The case $p_\mathrm{g.s.}$ is located in the interior of $\overline{p_7p_6}$, $\overline{p_7p_3}$ or $\overline{p_7p_8}$.}

When $p_\mathrm{g.s.}$ is in the interior of $\overline{p_7p_3}$, $S_A=N_A$, $S_B=N_B$, and $0<S_C<N_C$. The associated texture is $f$//$f$//$q$. The two inequalities $\frac{\partial E_\mathrm{spin}^\mathrm{M}}{\partial S_A}|_{p_\mathrm{g.s.}}<0$ and $\frac{\partial E_\mathrm{spin}^\mathrm{M}}{\partial S_B}|_{p_\mathrm{g.s.}}<0$, together with $\frac{\partial E_\mathrm{spin}^\mathrm{M}}{\partial S_C}|_{p_\mathrm{g.s.}}=0$ and $\frac{\partial^2E_\mathrm{spin}^\mathrm{M}}{\partial S_C^2}|_{p_\mathrm{g.s.}}>0$ are required. This leads to the constraint listed in the second row of Tab.\ref{tab1}. This texture can be realized only if $Q_C>0$ (i.e., the $C$-species is polar in nature), whereas $Q_A$ and $Q_B$ can be negative or weakly positive. If they are positive and large, the inter-species interaction should be even stronger to ensure that the inequalities hold. The equality for $S_C$ implies that the intra-force and the inter-force imposed on the $C$-atoms arrive at a balance. The energy $E_{ffq}^\mathrm{M}$ is given in Tab.\ref{tab2}. The textures $f$//$q$//$f$ and $q$//$f$//$f$ can be similarly discussed. These three together are called the double-$f$-texture (double-$f$-tex).

\subsubsection*{The case $p_\mathrm{g.s.}$ is located in the interior of the rectangles of the second kind.}

When $p_\mathrm{g.s.}$ is in the interior of $p_7p_6p_2p_3$, $S_A=N_A$, $0<S_B<N_C$, and $0<S_C<N_C$. The associated texture is $f$//$q$//$q$. The inequality $\frac{\partial E_\mathrm{spin}^\mathrm{M}}{\partial S_A}|_{p_\mathrm{g.s.}}<0$ together with $\frac{\partial E_\mathrm{spin}^\mathrm{M}}{\partial S_{J'}}|_{p_\mathrm{g.s.}}=0$ and $\frac{\partial^2E_\mathrm{spin}^\mathrm{M}}{\partial S_{J'}^2}|_{p_\mathrm{g.s.}}>0$ ($J'=B$ and $C$) are required. This leads to the constraint listed in the third row of Tab.\ref{tab1}. This texture can be realized only if both the $B$- and $C$-species are polar in nature, whereas $Q_A$ can be negative or weakly positive. Besides, the condition $Q_BQ_C>|Q_{BC}|^2$ is necessary. One can prove that the constraint listed in the third row leads to $\alpha_{ABC}<0$. Note that $E_{ppp}^\mathrm{M}=0$ while $E_{fqq}^\mathrm{M}$ is a product of a positive value and $\alpha_{ABC}$. Thus, $\alpha_{ABC}<0$ is a necessary condition for the $f$//$q$//$q$ texture. The textures $q$//$f$//$q$ and $q$//$q$//$f$ can be similarly discussed. The three together are called the single-$f$-tex.

\subsubsection*{The case $p_\mathrm{g.s.}$ is located at $p_1$.}

When all the three species are polar in nature ($Q_A>0$, $Q_B>0$, $Q_C>0$) and the inter-species forces are zero or weak, the first term of $\alpha_{ABC}$, $Q_AQ_BQ_C$, is positive and is dominant. This leads to $\alpha_{ABC}\geq 0$. In this case all the species are in $p$ and the texture is therefore denoted as $p$+$p$+$p$. When $\{|Q_{JJ_+}|\}$ increases, $\alpha_{ABC}$ will decrease. Once $\alpha_{ABC}$ becomes zero, the energy of the single-$f$-tex will be lower than $E_{ppp}^\mathrm{M}$ (refer to Tab.\ref{tab2}), and the transition from $p$+$p$+$p$ to the single-$f$-tex will occur.

With these in mind, the g.s. is either in the $p$+$p$+$p$ or in a texture without $p$ but with at least one species in $f$.

\subsection*{Spin-texture transition}

We aim at the effect caused by the variation of the inter-species forces. Note that the effect of $Q_{JJ_+}$ is to pull the spins of the $J$ and $J_+$ species lying along the same direction (opposite directions) if $Q_{JJ_+}<0$ ($>0$). Therefore, in general, a stronger $|Q_{JJ_+}|$ will cause the appearance of the $f$-phase. Starting from $\{|Q_{JJ_+}|\}=0$, the first transition is from $p$+$p$+$p$ to a single-$f$-tex as mentioned above. Recall that the single-$f$-tex must have $\alpha_{ABC}\leq 0$ while the $p$+$p$+$p$ has $\alpha_{ABC}>0$, therefore $\alpha_{ABC}=0$ is the critical point of transition. Since $\alpha_{ABC}$ is invariant under cyclic permutation of the indexes, this critical point is common to all $p$+$p$+$p\rightarrow$single-$f$-tex transitions disregarding which species is in $f$. One can prove that the three sets of constraint for the three single-$f$-texs do not compromise with each others, i.e., for a given set of parameters, the two sets of constraint for two different single-$f$-texs cannot both be satisfied. If both were satisfied, the combined constraints would lead to $\alpha_{ABC}>0$, and therefore in contradiction with the common feature $\alpha_{ABC}\leq 0$. This fact implies that, once a single-$f$-tex appears, the other two cannot appear. Thus, the transitions among the three single-$f$-texs (say, $f$//$q$//$q\rightarrow q$//$f$//$q$) are prohibited. Therefore, $p$+$p$+$p$ can transit only to a specific single-$f$-tex, it depends on the parameters.

When $\{|Q_{JJ_+}|\}$ increase further, a $q$-phase can be changed to an $f$-phase. Therefore, the single-$f$-tex$\rightarrow$double-$f$-tex transition will occur (as shown below). One can prove that the three sets of (in)equalities for the three double-$f$-texs do not compromise with each others as before. Thus, a single-$f$-tex can uniquely transit to a specific double-$f$-tex, and the transitions among the three double-$f$-texs are prohibited. When $\{|Q_{JJ_+}|\}$ increases further, eventually, the g.s. must be in the $f$//$f$//$f$ texture.

With these in mind the increase of $\{|Q_{JJ_+}|\}$ will lead to a chain of transitions as $p$+$p$+$p\rightarrow$single-$f$-tex$\rightarrow $double-$f$-tex$\rightarrow f$//$f$//$f$.

\begin{figure}[tbp]
 \centering \resizebox{0.6\columnwidth}{!}{\includegraphics{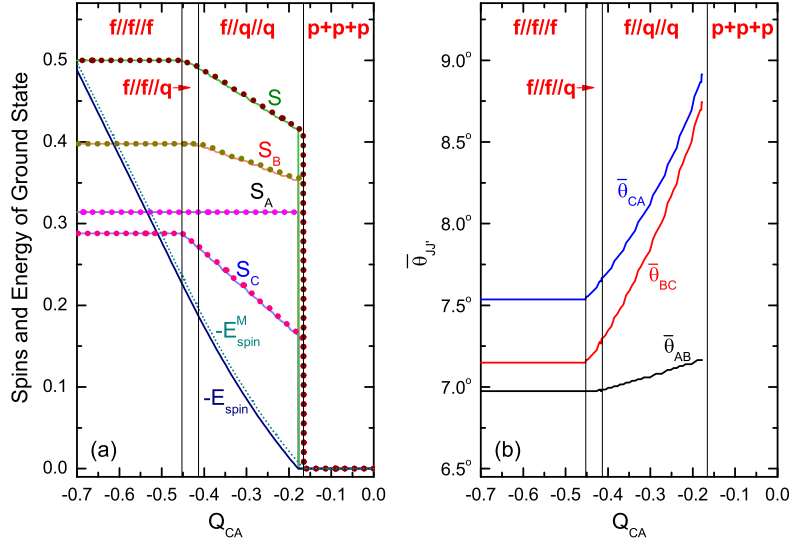} }
 \caption{(color online) An example for the variation of the spin-texture of Type-I against $Q_{CA}$. The texture is specified by $S_A/N$, $S_B/N$, $S_C/N$, and $S/(2N)$ (where $N=N_A+N_B+N_C)$ in (a) and by the angles $\bar{\theta}_{AB}$, $\bar{\theta}_{BC}$ and $\bar{\theta}_{CA}$ (in degree) between them (b). The results from the exact diagonalization of $H_\mathrm{spin}$ are plotted in solid lines. In (a), the results from the model are plotted in dotted lines, and $\theta_{AB}=\theta_{BC}=\theta_{CA}=0$ are assumed. Accordingly, the classical model has $S=S_{class}\equiv S_A+S_B+S_C$ as shown in (a). The dimensionless parameters are given as $N_A=120$, $N_B=152$, $N_C=110$, $Q_A=0.6$, $Q_B=0.5$, $Q_C=0.77$, $Q_{AB}=-0.46$, $Q_{BC}=-0.2$, $Q_{CA}$ is from $-0.7$ to $0$. Since all $\{Q_{JJ_+}\}$ are given negative, this example represents the case of Fig.\ref{fig2}a.}
 \label{fig4}
\end{figure}

\begin{figure}[tbp]
 \centering \resizebox{0.6\columnwidth}{!}{\includegraphics{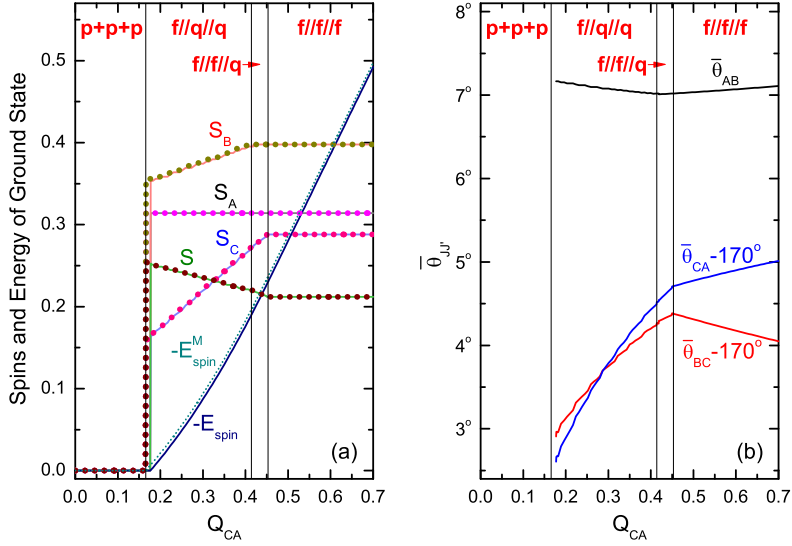} }
 \caption{(color online) An example similar to Fig.\ref{fig4} but with $Q_{BC}=0.2$ and $Q_{CA}$ is from $0$ to $0.7$. Since only one of $\{Q_{JJ_+}\}$ is given negative ($Q_{AB}=-0.46$), this example represents the case of Fig.\ref{fig2}b. Accordingly, in the model, $\theta_{AB}=0$ and $\theta_{BC}=\theta_{CA}=180^{\circ}$ are assumed and $S_{class}\equiv|S_A+S_B-S_C|$ in (a).}
 \label{fig5}
\end{figure}

Two numerical examples of Type I are shown in Fig.\ref{fig4} and Fig.\ref{fig5}, where the variation of the spin-texture (specified by $S_A$, $S_B $, $S_C$, and $\bar{\theta_{AB}}$, $\bar{\theta_{BC}}$, $\bar{\theta_{CA}}$) against $Q_{CA}$ is plotted. The results from the QM calculation are in solid lines, those from the model are in dotted lines. The coincidence is quite well. In particular, the whole chain of transitions is recovered by the QM calculation and the critical points are one-to-one close to each other. The intuitive pictures shown in Fig.\ref{fig2}a and Fig.\ref{fig2}b are also supported by Fig.\ref{fig4}b and Fig.\ref{fig5}b. In Fig.\ref{fig4}b the angles are very small $<9^{\circ}$), in Fig.\ref{fig5}b the angles are either close to zero or to $\pi$. Thus, the analysis based on the model is reliable. Note that the model is symmetric with respect to $Q_{JJ_+}\leftrightarrow -Q_{JJ_+}$. This symmetry can be shown by comparing Fig.\ref{fig4}a and Fig.\ref{fig5}a.

According to the model, when $|Q_{CA}|$ increases, the transition $p$+$p$+$p\rightarrow f$//$q$//$q$ occurs at $|Q_{CA}|=q_1$, where $E_{ppp}^\mathrm{M}=E_{fqq}^\mathrm{M}$. Thus, $q_1$ is the solution of the equation
\begin{equation}
 \alpha_{ABC}=0. \label{qabc}
\end{equation}
In Fig.\ref{fig4} $q_1=0.165$ as listed in Tab.\ref{tab3}. Recall that, for a 2-species BEC, the $p$+$p\rightarrow f$//$q$ transition will occur when $\alpha_{JJ_+}=0$ \cite{ref11,he1,he2}. Obviously, Eq.(\ref{qabc}) is a generalization of Eq.(\ref{abc}). In both equations the competition of the intra- and inter-interactions is clearly shown.

The transition $f$//$q$//$q\rightarrow f$//$f$//$q$ occurs at $q_2$, where $E_{fqq}^\mathrm{M}=E_{ffq}^\mathrm{M}$. Thus
\begin{equation}
 q_2
  =  \frac{1}{N_A|Q_{BC}|}
     ( N_BQ_BQ_C
      -N_B|Q_{BC}|^2
      -N_AQ_C|Q_{AB}| ). \label{q2}
\end{equation}
In Fig.\ref{fig4} $q_2=0.414$ as listed in Tab.\ref{tab3}.

The transition $f$//$f$//$q\rightarrow f$//$f$//$f$ occurs at $q_3,$ where $E_{ffq}^\mathrm{M}=E_{fff}^\mathrm{M}$. Thus,
\begin{equation}
 q_3
  =  \frac{1}{N_A}
     ( N_CQ_C
      -N_B|Q_{BC}|). \label{q3}
\end{equation}
In Fig.\ref{fig4} $q_3=0.453$. Recall that, for 2-species BEC with A and C atoms, the $f$//$q\rightarrow f$//$f$ transition will occur when $q_3=\frac{1}{N_A}N_CQ_C$ \cite{ref11,he1,he2}. Thus, the existence of the third species (B-atoms) is helpful to the transition (i.e., the $f$//$f$//$f$ texture can be realized at a smaller $|Q_{CA}|$).

It turns out that the critical values predicted by the model are close to the values from QM calculation as shown in Tab.\ref{tab3} (except $q_1$, but still acceptable). Thus, the related analytical formulae are useful for qualitative evaluation. For other chains of transition, the analytical formulae of the critical points can be similarly obtained.

\begin{table}[tbh]
 \caption{The critical values of $Q_{CA}$ in the chain $p$+$p$+$p\rightarrow f$//$q$//$q\rightarrow f$//$f$//$q\rightarrow f$//$f$//$f$. The other parameters are listed in the caption of Fig.\ref{fig5}.}
 \label{tab3}
 \begin{center}
 \begin{tabular}{c|c|c}
 \hline\hline
 critical point & classical model       & QM calculation    \\ \hline
 $q_1$          & $0.165$               & $0.177$           \\
 $q_2$          & $0.414$               & $0.419$           \\
 $q_3$          & $0.453$               & $0.450$           \\
 \hline\hline
 \end{tabular}
 \end{center}
\end{table}

\subsection*{Classical model (Type-II)}

When all $Q_{JJ_+}$ are positive (Fig.\ref{fig2}c) or only one of them is positive (Fig.\ref{fig2}d, where $Q_{AB}>0$), the associated spin-textures are in Type-II. In this type the three spins point at different directions, but they are assumed to be coplanar ($\theta_{AB}+\theta_{BC}+\theta_{CA}=2\pi$). The total energy appears as
\begin{equation}
 E_\mathrm{spin}^\mathrm{M}
  =  \sum_J
     Q_J
     S_J^2
    +2\sum_J
     Q_{JJ_+}'
     S_J
     S_{J_+}, \label{emt1p}
\end{equation}
where $Q_{JJ_+}'=Q_{JJ_+}\cos \theta_{JJ_+}$.

To find out the point $p_\mathrm{g.s.}$ where the minimum of $E_\mathrm{spin}^\mathrm{M}$ is located, we first consider the partial derivatives of $E_\mathrm{spin}^\mathrm{M}$ against $\{S_J\}$ when $\{Q_J\}$ and $\{Q_{JJ_+}'\}$ are considered as constants. Thus, the situation is the same as for Type-I. With the same arguments as those for Type-I, we deduce that $p_\mathrm{g.s.}$ can only access $p_1$, the interiors of $\overline{p_7p_6}$, $\overline{p_7p_3}$, $\overline{p_7p_8}$, the interiors of the three rectangles of the second kind, and $p_7$.

\subsubsection*{$p_\mathrm{g.s.}=p_7$.}

In this case every species is fully polarized, but the spins of any two species are in general neither parallel nor antiparallel to each other. Therefore, instead of $f$//$f$//$f$, this type of texture is denoted as $f$+$f$+$f$. The three inequalities $\frac{\partial E_\mathrm{spin}^\mathrm{M}}{\partial S_J}|_{p_\mathrm{g.s.}}<0$ are required which lead to the constraints $N_JQ_J+N_{J_-}Q_{J_-J}'+N_{J_+}Q_{JJ_+}'<0$, where $J$ is for $A$, $B$ and $C$. In addition, the two derivatives $\frac{\partial E_\mathrm{spin}^\mathrm{M}}{\partial \theta_{BC}}|_{p_\mathrm{g.s.}}$ and $\frac{\partial E_\mathrm{spin}^\mathrm{M}}{\partial \theta_{CA}}|_{p_\mathrm{g.s.}}$ are required to be zero. These lead to (refer to Eqs.(\ref{em3}) and (\ref{em4}))
\begin{eqnarray}
 \cos\theta_{BC}
 &=& G_{BC}(N_{A}N_{B}N_{C}),  \label{cos1}  \\
 \cos\theta_{CA}
 &=& G_{CA}(N_{A}N_{B}N_{C}).  \label{cos2}
\end{eqnarray}
The two angles obtained in this way should ensure that the two second order derivatives given in Eqs.(\ref{em1p}) and (\ref{em2p}) are positive. When all the $Q_{JJ_+}>0$, from Eqs.(\ref{em1p}) and (\ref{em2p}) we know that this requirement could be satisfied if $\theta_{BC}$ and $\theta_{CA}$ are large enough, thereby the repulsion caused by $Q_{JJ_+}$ is reduced. Whereas when only one, say, $Q_{AB}>0$ while $Q_{BC}<0$ and $Q_{CA}<0$, $\theta_{BC}$ and $\theta_{CA}$ should be small enough, thereby the attraction caused by $Q_{BC}$ and $Q_{CA}$ can be strengthened. Once the angles are known, the three $Q_{JJ'}'$ are known. Then, the energy $E_\mathrm{spin}^\mathrm{M}=\sum_JQ_JN_J^2+2\sum_JQ_{JJ_+}'N_JN_{J_+}\equiv E_{f+f+f}$ and the subspace of parameters that supports this texture are also known.

\subsubsection*{$p_\mathrm{g.s.}$ locates in the interiors of $\overline{p_7p_6}$, $\overline{p_7p_3}$ and $\overline{p_7p_8}$.}

When $p_\mathrm{g.s.}$ locates in the interior of $\overline{p_7p_3}$ as an example, $S_A=N_A$, $S_B=N_B$, and the texture is denoted as $f$+$f$+$q$. The constraints appear as (refer to the second row of Tab.\ref{tab1}):
\begin{eqnarray}
 \left\{
 \begin{array}{l}
   N_AQ_A+N_BQ_{AB}'+S_CQ_{CA}'<0 \\
   N_BQ_B+S_CQ_{BC}'+N_AQ_{AB}'<0 \\
   S_CQ_C+(N_AQ_{CA}'+N_BQ_{BC}')=0
 \end{array}
 \right., \label{ffq2}
\end{eqnarray}
The angles are subjected to the two coupled equations (refer to Eqs.(\ref{em3}) and (\ref{em4}))
\begin{eqnarray}
 \cos\theta_{BC}
 &=& G_{BC}(N_A,N_B,-(N_AQ_{CA}'+N_BQ_{BC}')/Q_C), \label{ctbcp} \\
 \cos\theta_{CA}
 &=& G_{CA}(N_A,N_B,-(N_AQ_{CA}'+N_BQ_{BC}')/Q_C),  \label{ctcap}
\end{eqnarray}
where the angles are also contained in $Q_{JJ_+}'$. Solving these equations (say, numerically), we can obtain $\theta_{BC}$ and $\theta_{CA}$. Then, the energy $E_{f+f+q}$ and the subspace of parameters that supports this texture can be known as before. The cases of $f$+$q$+$f$ and $q$+$f$+$f$ can be similarly discussed.

\subsubsection*{$p_\mathrm{g.s.}$ locates in the interiors of the rectangles of the second kind.}

For the rectangle $p_7p_6p_2p_3$ as an example, $S_A=N_A$, the texture is denoted as $f$+$q$+$q$. The constraint imposed on this texture is listed in the third row of Tab.\ref{tab1} but with $-|Q_{JJ_+}|$ being replaced by $Q_{JJ}'$. In addition, the two coupled equations
\begin{eqnarray}
 \cos\theta_{BC}
 &=& G_{BC}(N_AS_BS_C), \label{ctbc2} \\
 \cos\theta_{CA}
 &=& G_{CA}(N_AS_BS_C), \label{ctca2}
\end{eqnarray}
are required to be satisfied. Then $S_B$, $S_C$, together with the angles can be known, thereby $E_{f+q+q}^\mathrm{M}$ is known.

\subsubsection*{$p_\mathrm{g.s.}=p_1$.}

When all $Q_J>0$ and when the strengths of the inter-species interaction become weaker, all the three $E_{f+q+q}^\mathrm{M}$, $E_{q+f+q}^\mathrm{M}$, and $E_{q+q+f}^\mathrm{M}$ will be larger than zero, in this case $p_\mathrm{g.s.}=p_1$ and the texture is $p$+$p$+$p$.

A comparison of the results from the model and from the diagonalization of $H_\mathrm{spin}$ is shown in Tab.\ref{tab4}.

\begin{table}[tbh]
 \caption{For the texture $f$+$f$+$f$ of the Type-II., the angles (in degrees) between the spins against the increase of $Q_{CA}$. The data for $\theta_{JJ_+}$ are from the model (refer to Eqs.(\ref{cos1}) and (\ref{cos2})), those for $\bar{\theta}_{JJ_+}$ are from the diagonalization of $H_\mathrm{spin}$ (refer to Eqs.(\ref{ctab}), (\ref{ctbc}) and (\ref{ctca})). The parameters are given as $N_A=120$, $N_B=152$, $N_C=110$, $Q_A=-0.6$, $Q_B=-0.5$, $Q_C=-0.77$, $Q_{AB}=0.3$, $Q_{BC}=0.4$, $Q_{CA}$ is from $0.3$ to $0.8$.}
 \label{tab4}
 \begin{center}
 \begin{tabular}{c|rr|cc|rr|c}
 \hline\hline
 $Q_{CA}$ &
 $\theta_{CA}$ &
 $\bar{\theta}_{CA}$ &
 $\theta_{BC}$ &
 $\bar{\theta}_{BC}$ &
 $\theta_{AB}$ &
 $\bar{\theta}_{AB}$ &
 $\bar{\theta}_{CA}+\bar{\theta}_{BC}+\bar{\theta}_{AB}$ \\ \hline
 $0.3$  & $ 81.6$       & $ 81.8$       & $144.2$       & $144.0$       & $134.3$       & $134.1$       & $359.9$       \\
 $0.4$  & $111.3$       & $111.1$       & $132.7$       & $132.5$       & $116.0$       & $116.4$       & $360.0$       \\
 $0.5$  & $126.9$       & $126.7$       & $127.9$       & $127.6$       & $105.3$       & $105.5$       & $359.8$       \\
 $0.6$  & $136.8$       & $136.7$       & $125.8$       & $125.5$       & $ 97.5$       & $ 97.6$       & $359.8$       \\
 $0.7$  & $143.7$       & $143.5$       & $125.1$       & $124.7$       & $ 91.2$       & $ 91.6$       & $359.8$       \\
 $0.8$  & $148.9$       & $148.5$       & $125.3$       & $124.6$       & $ 85.8$       & $ 86.6$       & $359.7$       \\
 \hline\hline
 \end{tabular}
 \end{center}
\end{table}

Tab.\ref{tab4} demonstrates that the results given by Eqs.(\ref{ctab}), (\ref{ctbc}) and (\ref{ctca}) are quite accurate. In particular, the sum of the three $\{\bar{\theta}_{JJ_+}\}$ given in the last column is very close to $2\pi$. This supports the assumption of coplanar texture.

\section*{Final remarks}

Features of the spin-textures of 3-species condensates with spin-1 atoms have been extracted from a model and have been checked via QM calculation. The results from the model are found to be valid. In summary:

\begin{itemize}
\item The textures can be described by the norms of the three spins $\{S_J\}$ and the average angles $\{\bar{\theta}_{JJ_+}\}$ between them. When the three species are polar in nature (i.e., all $c_{J2}>0$) and the inter-forces are weak, $\{S_J\}$ can all be zero ($p$+$p$+$p$). Otherwise, they are all nonzero and essentially lying on a plane.

\item The spin-textures not in $p$+$p$+$p$ can be first classified according to the relative orientations of $\{S_J\}$ as intuitively shown in Fig.\ref{fig2}. When all inter-forces are attractive (i.e., all $c_{JJ'2}<0$), the texture is shown in Fig.\ref{fig2}a where all spins point to the same direction. When only one is attractive (say, $c_{AB2}<0$, ), shown in Fig.\ref{fig2}b. When all are repulsive (all $c_{JJ'2}>0$), shown in Fig.\ref{fig2}c. When only one is repulsive (say, $c_{AB2}>0$), shown in Fig.\ref{fig2}d.

\item The spin-textures can be further classified according to the norms of the spin. In addition to $p$+$p$+$p$, there are other three textures, namely, the single-$f$-tex (where one species is in $f$, i.e., fully polarized), the double-$f$-tex (two species in $f$), and the $f$+$f$+$f$ (all in $f$). Note that the single-$p$-tex, the double-$p$-tex, and the $q$+$q$+$q$ do not exist. Thus, the coexistence of a $p$ and an $f$ (or a $q$) is not allowed. If not in $p$+$p$+$p$, at least a species must be fully polarized.

\item Starting from the $p$+$p$+$p$, when $|c_{JJ'2}|$ increases, more species will tend to be in $f$-phase. Therefore, a chain of phase-transitions $p$+$p$+$p\rightarrow f$+$q$+$q\rightarrow f$+$f$+$q\rightarrow f$+$f$+$f$ will occur. In the parameter space, there are a number of critical surfaces. When the point (representing a set of parameters) vary and pass through one of the surfaces, a transition will occur. For Type-I (Fig.\ref{fig2}a and Fig.\ref{fig2}b) the equations describing the surfaces have been quite accurately obtained (refer to Eqs.(\ref{qabc}), (\ref{q2}) and (\ref{q3})). Thus, the critical points at which the transitions occur can be predicted. Moreover, the analytical formulae demonstrate the competition among contradicting physical factors, thereby the inherent physics could be understood better. For Type-II (Fig.\ref{fig2}c and Fig.\ref{fig2}d), analytical analysis based on the model becomes complicated. Nonetheless, the results from the model have been checked to be also valid.
\end{itemize}

\section*{Acknowledgements}

Supported by the National Natural Science Foundation of China under Grants No.11372122, 11274393, 11574404, and 11275279; the Open Project Program of State Key Laboratory of Theoretical Physics, Institute of Theoretical Physics, Chinese Academy of Sciences, China(No.Y4KF201CJ1); the National Basic Research Program of China (2013CB933601); and the Natural Science Foundation of Guangdong of China (2016A030313313).

\section*{Appendix}

From the total energy of the model given in Eq.(\ref{em}) we have the derivatives:
\begin{eqnarray}
 \frac{\partial E_\mathrm{spin}^\mathrm{M}}{\partial\theta_{BC}}
 &=&-2[ S_AS_BQ_{AB}
        \sin(\theta_{BC}+\theta_{CA})
       +S_BS_CQ_{BC}
        \sin\theta_{BC} ], \label{em1} \\
 \frac{\partial E_\mathrm{spin}^\mathrm{M}}{\partial\theta_{CA}}
 &=&-2[ S_AS_BQ_{AB}
        \sin(\theta_{BC}+\theta_{CA})
       +S_CS_AQ_{CA}
        \sin\theta_{CA} ], \label{em2} \\
 \frac{\partial^2E_\mathrm{spin}^\mathrm{M}}{\partial\theta_{BC}^2}
 &=&-2[ S_AS_BQ_{AB}
        \cos(\theta_{BC}+\theta_{CA})
       +S_BS_CQ_{BC}
        \cos\theta_{BC} ], \label{em1p} \\
 \frac{\partial^2E_\mathrm{spin}^\mathrm{M}}{\partial\theta_{CA}^2}
 &=&-2[ S_AS_BQ_{AB}
        \cos(\theta_{BC}+\theta_{CA})
       +S_CS_AQ_{CA}
        \cos\theta_{CA}]. \label{em2p}
\end{eqnarray}
Note that the coupled equations $\frac{\partial E_\mathrm{spin}^\mathrm{M}}{\partial\theta_{BC}}=0$ and $\frac{\partial E_\mathrm{spin}^\mathrm{M}}{\partial\theta_{CA}}=0$ have a trivial solution: both $\theta_{BC}$ and $\theta_{CA}$ are equal to $0$ or $\pi$, and a non-trivial solution as
\begin{eqnarray}
 \cos\theta_{BC}
 &=& G_{BC}(S_AS_BS_C)
 \equiv
     \frac{ (S_AQ_{AB}Q_{CA})^2
           -(S_BQ_{AB}Q_{BC})^2
           -(S_CQ_{BC}Q_{CA})^2}
          {2S_BS_CQ_{AB}Q_{BC}^2Q_{CA}}, \label{em3} \\
 \cos\theta_{CA}
 &=& G_{CA}(S_AS_BS_C)
 \equiv
     \frac{-(S_AQ_{AB}Q_{CA})^2
           +(S_BQ_{AB}Q_{BC})^2
           -(S_CQ_{BC}Q_{CA})^2}
          {2S_AS_CQ_{AB}Q_{BC}Q_{CA}^2}. \label{em4}
\end{eqnarray}
Besides, one can prove the following useful relation
\begin{equation}
 \sin\theta_{CA}
  =  \frac{S_BQ_{BC}}{S_AQ_{CA}}
     \sin\theta_{BC}. \label{cabc}
\end{equation}
We further have
\begin{eqnarray}
 \frac{\partial E_\mathrm{spin}^\mathrm{M}}{\partial S_J}
 &=& 2[ Q_JS_J
       +Q_{J_-J}
        \cos\theta_{J_-J}
        S_{J_-}
       +Q_{JJ_+}
        \cos\theta_{JJ_+}
        S_{J_+} ],\ \ \
        (J=A,B,C) \label{em5} \\
 \frac{\partial^2E_\mathrm{spin}^\mathrm{M}}{\partial S_J^2}
 &=& 2Q_J. \label{em5p}
\end{eqnarray}
The above partial derivatives of $E_\mathrm{spin}^\mathrm{M}$ are essential in the search of the g.s..

\section*{Author contributions}

Y. Z. He is responsible to the numerical calculation. Y. M. Liu is responsible to the theoretical derivation. C. G. Bao provides the idea, write the paper, and responsible to the whole paper. All authors reviewed the manuscript.

\section*{Additional information}

\textbf{Competing Interests:} The authors declare that they have no competing interests.


\begin{thebibliography}{99}
\bibitem{ref5}
 Stamper-Kurn, D. M., Andrews, M. R., Chikkatur, A. P., Inouye, S., Miesner, H. J., Stenger, J. and Ketterle, W.,
 Optical Confinement of a Bose-Einstein Condensate,
 \textit{Phys. Rev. Lett.} \textbf{80}, 2027 (1998).

\bibitem{ref6}
 Ho, T. L.,
 Spinor Bose Condensates in Optical Traps,
 \textit{Phys. Rev. Lett.} \textbf{81}, 742 (1998).

\bibitem{ref7}
 Law, C. K., Pu, H. and Bigelow, N. P.,
 Quantum Spins Mixing in Spinor Bose-Einstein Condensates,
 \textit{Phys. Rev. Lett.} \textbf{81}, 5257 (1998).

\bibitem{ref8}
 Goldstein, Elena V. and Meystre, Pierre,
 Quantum theory of atomic four-wave mixing in Bose-Einstein condensates,
 \textit{Phys. Rev. A} \textbf{59}, 3896 (1999).

\bibitem{ref9}
 Ho, T. L. and Yip, S. K.,
 Fragmented and Single Condensate Ground States of Spin-1 Bose Gas,
 \textit{Phys. Rev. Lett.} \textbf{84}, 4031 (2000).

\bibitem{ref10}
 Koashi, M. and Ueda, M.,
 Exact Eigenstates and Magnetic Response of Spin-1 and Spin-2 Bose-Einstein Condensates,
 \textit{Phys. Rev. Lett.} \textbf{84}, 1066 (2000).

\bibitem{ml}
 Luo, M., Li, Z. B. and Bao, C. G.,
 Bose-Einstein condensate of a mixture of two species of spin-1 atoms,
 \textit{Phys. Rev. A} \textbf{75}, 043609 (2007).

\bibitem{ref13}
 Xu, Z. F., L\"{u}, R. and You, L.,
 Quantum entangled ground states of two spinor Bose-Einstein condensates,
 \textit{Phys. Rev. A} \textbf{84}, 063634 (2011).

\bibitem{ref14}
 Shi, Yu and Ge, Li,
 Three-dimensional quantum phase diagram of the exact ground states of a mixture of two species of spin-1 Bose gases with interspecies spin exchange,
 \textit{Phys. Rev. A} \textbf{83}, 013616 (2011).

\bibitem{ref15}
 Shi, Yu and Ge, Li,
 Ground states of a mixture of two species of spin-1 Bose gases with interspecies spin exchange in a magnetic field,
 \textit{Int. J. Mod. Phys. B} \textbf{26}, 1250002 (2012).

\bibitem{ref16}
 Xu, Z. F., Mei, J. W., L\"{u}, R. and You, L.,
 Spontaneously axisymmetry-breaking phase in a binary mixture of spinor Bose-Einstein condensates,
 \textit{Phys. Rev. A} \textbf{82}, 053626 (2010).

\bibitem{ref17}
 Zhang, J., Li, T. T. and Zhang, Yunbo,
 Interspecies singlet pairing in a mixture of two spin-1 Bose condensates,
 \textit{Phys. Rev. A} \textbf{83}, 023614 (2011).

\bibitem{ref18}
 Irikura, Naoki, Eto, Yujiro, Hirano, Takuya and Saito, Hiroki,
 Ground-state phases of a mixture of spin-1 and spin-2 Bose-Einstein condensates,
 \textit{Phys. Rev. A} \textbf{97}, 023622 (2018).

\bibitem{ref12}
 Shi, Yu,
 Ground states of a mixture of two species of spinor Bose gases with interspecies spin exchange,
 \textit{Phys. Rev. A} \textbf{82}, 023603 (2010).

\bibitem{he1}
 He, Y. Z., Liu, Y. M. and Bao, C. G.,
 Variation of the spin textures of 2-species spin-1 condensates studied beyond the single spatial mode approximation and the experimental identification of these textures,
 \textit{Phys. Scr.} \textbf{94}, 115403 (2019).

\bibitem{ref11}
 Xu, Z. F., Zhang, Yunbo and You, L.,
 Binary mixture of spinor atomic Bose-Einstein condensates,
 \textit{Phys. Rev. A} \textbf{79}, 023613 (2009).

\bibitem{he2}
 He, Y. Z., Liu, Y. M. and Bao, C. G.,
 Spin-Textures of the Condensates with Two Kinds of Spin-1 Atoms Studied Beyond the Single Spatial Mode Approximation,
 \textit{J. Low Temp. Phys.} \textbf{196}, 458-472 (2019).

\bibitem{katr}
 Katriel, J.,
 Weights of the total spins for systems of permutational symmetry adapted spin-1 particles,
 \textit{Journal of Molecular Structure: THEOCHEM} \textbf{547}, 1-11 (2001).

\bibitem{polo}
 J.Polo, \textit{et al.},
 Analysis beyond the Thomas-Fermi approximation of the density profiles of a miscible two-component Bose-Einstein condensate,
 \textit{Phys. Rev. A }\textbf{91}, 053626 (2015).

\bibitem{he}
 He, Y. Z., Liu, Y. M. and Bao, C. G.,
 Generalized Gross-Pitaevskii equation adapted to the U(5)$\supset$SO(5)$\supset$SO(3) symmetry for spin-2 condensates,
 \textit{Phys. Rev. A} \textbf{91}, 033620 (2015).

\bibitem{sr1}
 Liu, Y. M., He, Y. Z. and Bao, C. G.,
 Singularity in the matrix of the coupled Gross-Pitaevskii equations and the related state-transitions in three-species condensates,
 \textit{Scientific reports} \textbf{7}, 6585 (2017).

\bibitem{zbl}
 Li, Z. B., Liu, Y. M., Yao, D. X. and Bao, C. G.,
 Two types of phase diagrams for two-species Bose-Einstein condensates and the combined effect of the parameters,
 \textit{J. Phys. B: At. Mol. Opt. Phys.} \textbf{50}, 135301 (2017).

\bibitem{zbl2}
 Bao, C. G. and Li, Z. B.,
 Ground band and a generalized Gross-Pitaevskii equation for spinor Bose-Einstein condensates,
 \textit{Phys. Rev A} \textbf{70}, 043620 (2004).
\end{thebibliography}
\end{document}